\documentstyle [12pt] {article}

\newcommand{\gtwid}{\mathrel{\raise.3ex\hbox{$>$\kern-.75em\lower1ex
\hbox{$\sim$}}}}
\newcommand{\ltwid}{\mathrel{\raise.3ex\hbox{$<$\kern-.75em\lower1ex
\hbox{$\sim$}}}}
\newcommand{\beq}{\begin{equation}}
\newcommand{\eeq}{\end{equation}}
\newcommand{\beqs}{\begin{eqnarray}}
\newcommand{\eeqs}{\end{eqnarray}}

\catcode`@=11
\@addtoreset{equation}{section}
\@addtoreset{equation}{subsection}
\def\theequation{\ifnum\value{section}=0 \arabic{equation}\ignorespaces
\else \ifnum\value{section}=-1 A.\arabic{equation}\ignorespaces 
\else \ifnum\value{subsection}=0 \thesection.\arabic{equation}\ignorespaces
\else \thesection.\arabic{subsection}.\arabic{equation}\ignorespaces
                           \fi
                      \fi
                 \fi}
\catcode`@=12

\begin{document}

\def\thefootnote{\fnsymbol{footnote}}
\baselineskip 6.0mm

\begin{flushright}
\begin{tabular}{l}
ITP-SB-94-53    \\
November, 1994  
\end{tabular}
\end{flushright}

\vspace{8mm}
\begin{center}
{\Large \bf Complex-Temperature Singularities in the}
\vspace{3mm}
{\Large \bf $d=2$ Ising Model. II. Triangular Lattice} 
\vspace{4mm}
\vspace{16mm}

\setcounter{footnote}{0}
Victor Matveev\footnote{email: vmatveev@max.physics.sunysb.edu}
\setcounter{footnote}{6}
and Robert Shrock\footnote{email: shrock@max.physics.sunysb.edu}

\vspace{6mm}
Institute for Theoretical Physics  \\
State University of New York       \\
Stony Brook, N. Y. 11794-3840  \\

\vspace{20mm}

{\bf Abstract}
\end{center}
   We investigate complex-temperature singularities in the Ising model on 
the triangular lattice.  Extending an earlier analysis of the low-temperature
series expansions for the (zero-field) susceptibility $\bar\chi$ by Guttmann
\cite{g75} to include the use of differential approximants, we obtain further 
evidence in support of his conclusion that the exponent describing the 
divergence in $\chi$ at $u=u_e=-1/3$ (where $u = e^{-4K}$) is 
$\gamma_e'=5/4$ and refine his estimate of the critical amplitude.  We discuss
the remarkable nature of this singularity, at which the spontaneous
magnetisation diverges (with exponent $\beta_e=-1/8$) and show that it lies at
the endpoint of a singular line segment constituting part of the natural
boundaries of the free energy in the complex $u$ plane. 
Using exact results, we find that the specific heat has a divergent 
singularity at $u=-1/3$ with exponent $\alpha_e'=1$, so that the relation 
$\alpha_e'+2\beta_e+\gamma_e'=2$ is satisfied.  We also study the singularity
at $u=u_s=-1$, where $M$ vanishes (with $\beta_s=3/8$) and $C$ diverges
logarithmically (with $\alpha_s' = \alpha_s = 0$). 
\vspace{35mm}

\pagestyle{empty}
\newpage

\pagestyle{plain}
\pagenumbering{arabic}
\renewcommand{\thefootnote}{\arabic{footnote}}
\setcounter{footnote}{0}

\section{Introduction}
\label{intro}

    Although exact closed-form expressions for the (zero-field) free energy 
and spontaneous magnetisation of the two-dimensional Ising model have 
been calculated, no such expression is known for the susceptibility. One can 
have at least two views on this.  The first is that since the problem has been
unsolved for 50 years, it is too difficult and one should consider other
problems more amenable to solution.  A second view is that some progress on 
even longstanding problems can be made with the input of new information.  In
support of this view, one could cite, for example, the recent progress
\cite{aw} on proving Fermat's last theorem, which has been unsolved for
more than 350 years since its first assertion in 1637. 
We shall adopt the second view here.  In this context, 
every additional piece of information on the susceptibility is 
of value, especially insofar as it specifies properties which an exact
solution must satisfy.  In particular, it is of interest to study the
susceptibility as an analytic function in the complex temperature plane. 
Complex-temperature singularities of the susceptibility have been known 
for many years.  When early low-temperature series expansions were calculated 
for the Ising model by the King's College group \cite{sykes1}, it was found 
that for the simple cubic, body-centered cubic, and face-centered cubic 
lattices, these have complex-temperature singularities closer to the origin 
than the position of the physical critical point, which precluded the use of 
a ratio test to determine the location of this physical critical point 
\cite{g69,dg}. For the triangular lattice, it was found that the series
expansions showed a complex-temperature singularity at the point \cite{g69,dg}
\beq
u = u_e = -\frac{1}{3}
\label{ue}
\eeq 
on the opposite side of the origin, and an equal distance from it, compared to
the physical critical point $u_c = 1/3$ for this lattice, where 
\beq
u = e^{-4K}
\label{udef}
\eeq
and $K$ denotes the spin-spin coupling (see below for notation). 
Transformations of the series variables to map the unphysical 
singularities farther away from the origin than the physical
critical point and methods of analysis such as dlog Pad\'{e} approximants
enabled one to obtain useful information on critical behaviour from
low-temperature series (for an early review, see Ref. \cite{gg}).  In
particular, as part of a general analysis of the susceptibility of the Ising 
model on 2D lattices \cite{g75} using high- and low-temperature series, 
Guttmann studied the complex-temperature singularity at $u=-1/3$ for the 
triangular lattice.  To obtain a series with good sensitivity to this
singularity, he subtracted the dominant physical singularity and its leading
confluent correction, and analysed the resultant series using ratio and dlog
Pad\'{e} methods.  From this analysis, Guttmann concluded that as $u$
approaches $u_e$ from the direction of the origin, the susceptibility has a 
divergent singularity of the form 
\beq
\bar\chi \simeq A_e'(1-u/u_e)^{-\gamma_e'}
\label{singform}
\eeq
with the exponent  
\beq
\gamma_e'= \frac{5}{4}
\label{gammae}
\eeq
and the critical amplitude 
\beq
A_e' = -0.0568 \pm 0.0008
\label{aeg}
\eeq

     Several years ago, we reported some results on complex-temperature
singularities of the susceptibility and correlation length 
for the Ising model \cite{ms}, including a proof that for
the square lattice, the (zero-field) susceptibility can have at most finite
non-analyticities on the border of the complex-temperature extension of the
symmetric phase, apart from its divergence at the physical critical
point. Before proceeding, we mention some related work.  Ref. \cite{ipz}
contains some discussion of complex-temperature singularities of the Ising
model on various 3D lattices. 
In a calculation and analysis of low-temperature series for the spin $1/2$
and spin 1 Ising models on the square lattice, Guttmann noted that for the 
spin $1/2$ model $\chi$ has a complex-temperature singularity consistent with 
the form $(1+u)^{-\gamma_s'}$, with $\gamma_s'=3/2$ (and $u=e^{-4K}$) 
\cite{egj}.  This paper also reported results for complex-temperature
singularities for the spin 1 Ising model on the square lattice. 
Slightly later, the present authors (unaware of Ref. \cite{egj}) carried out 
a similar study of the low-temperature series, inferred the same singularity 
in $\chi$ at $u=-1$ and value of the exponent $\gamma_s'$ and calculated the 
critical amplitude \cite{chisq}.  One of the results of Ref. \cite{chisq} of 
greatest potential value for constructing conjectures for the exact 
susceptibility 
was a unified treatment (c.f. eqs. (3.3.5)-(3.3.9)) of both the critical
singularity at $u=3-2\sqrt{2}$ and the singularity at $u=-1$ which yielded 
the relation
\beq
\gamma_s'=2(\gamma' - 1)
\label{gamrel}
\eeq 
where $\gamma'=7/4$ is the exponent at the physical critical point, as
approached from the FM phase.  Ref. \cite{chisq} also showed, using exact 
results, that
universality, in the sense that critical exponents are independent of lattice
type, is violated at complex-temperature singular points and provided an 
explanation of this as being related to the fact that the Hamiltonian is not,
in general, a real number at such points.  This work also showed that 
scaling and hyperscaling relations are not, in general, satisfied at
complex-temperature singular points.  In Ref. \cite{chisq} we emphasized the 
fact that, in accordance with the theorem of Ref. \cite{ms}, $\chi$ is 
finite (although non-analytic) as one approaches the point $u=-1$ from within 
the complex-temperature extension of the symmetric, paramagnetic (PM) phase, 
whereas it is divergent when one approaches this point from within the 
complex-temperature extension of the ferromagnetic (FM) or antiferromagnetic 
(AFM) phases, which behaviour is unprecedented for physical critical points. 
Hereafter, we shall denote Ref. \cite{chisq} as I.  In the present paper, we
shall continue the study of complex-temperature singular points of the Ising
model, dealing with the triangular lattice.  A companion paper treats the
closely related honeycomb lattice \cite{chihc}. 

\section{Complex Extensions of Physical Phases}
\label{general}

   In this section we define the model and our notation, which follows that in
paper I. We consider the Ising model on the triangular lattice (coordination
number $q=6$) at a temperature $T$ and external magnetic field $H$ defined 
in standard notation by the partition function 
$Z = \sum_{\{\sigma_i\}} e^{-\beta {\cal H}}$ with the Hamiltonian 
\beq
{\cal H} = -J \sum_{<ij>} \sigma_i \sigma_j - H \sum_i \sigma_i 
\label{h}
\eeq
where $\sigma_i = \pm 1$ are the $Z_2$ spin variables on each site $i$ of the
lattice, $\beta = (k_BT)^{-1}$, $J$ is the exchange constant, 
$<ij>$ denote nearest-neighbor pairs, and the magnetic moment $\mu \equiv
1$. We use the standard notation $K = \beta J$, $h = \beta H$, $u$ as defined
above in eq. (\ref{udef}), and 
\beq
v = \tanh K
\label{v}
\eeq
\beq
z = e^{-2K}
\label{u}
\eeq
which are related by the bilinear conformal mappings
\beq
z = \frac{1-v}{1+v} \ , \qquad v = \frac{1-z}{1+z} 
\label{bilinearmap}
\eeq
The reduced free energy per site is $f = \beta F = -N_s^{-1} \lim_{N_s \to 
\infty} \ln Z$ (where $N_s$ denotes the number of sites on the lattice), 
and the zero-field susceptibility is 
$\chi_0 =\frac{\partial M(H)}{\partial H}|_{H=0}$, where $M(H)$ denotes the 
magnetisation.  Henceforth, unless otherwise stated, we
consider only the case of zero external field and drop the subscript on
$\chi_0$.  It is convenient to define the related quantity $\bar\chi \equiv 
\beta^{-1}\chi$. After the pioneering solution for $f(K,h=0)$ on the square
lattice by Onsager \cite{ons}, several papers obtained solutions for $f(K,h=0)$
on the triangular lattice \cite{ftri}.  For reviews of the Ising model on the 
triangular lattice, see Refs. \cite{domb1,domb2,baxter}.  After Yang's
derivation of the spontaneous magnetisation $M$ for the square lattice
\cite{yang}, Potts calculated $M$ for the triangular lattice \cite{potts52}. 

We denote the critical coupling separating the paramagnetic, $Z_2$-symmetric 
high temperature phase from the phase with spontaneously broken $Z_2$ 
symmetry and ferromagnetic long-range order as $K_c$.  As was noted above, 
$u_c=1/3$, or equivalently, $z_c=1/\sqrt{3}$,  and 
$v_c=2-\sqrt{3}=0.267949...$.  As is well known, the Ising model on the 
triangular lattice (with equal exchange couplings along the three lattice 
vectors) does not have any phase transition to a phase with antiferromagnetic 
long-range order.  Indeed, even at $T=0$ (corresponding to $v=-1$ and 
$K = z = u = \infty$), there is no AFM long-range order \cite{steph1}. 

   Since we shall study the susceptibility as a function of complex (inverse) 
temperature, it is useful to discuss the complex-temperature extensions of the
physical PM and FM phases. 
As discussed in Ref. \cite{ms}, for zero external field, 
there is an infinite periodicity in complex $K$ under certain shifts along the
imaginary $K$ axis as a consequence of the fact that the spin-spin interaction
$\sigma_i\sigma_j$ in ${\cal H}$ is an integer.  In particular, there is an
infinite repetition of phases as functions of complex $K$; this infinitely
repeated set of phases is reduced to a single set by using the variables 
$v$, $z$ and/or $u$, owing to the fact that these latter variables have very 
simple properties under complex shifts in $K$: 
\beq
K \to K+ n i \pi \Rightarrow \{ v \to v, \ \ z \to z, \ \ u \to u\}
\label{shift1}
\eeq
\beq
K \to K+(n + \frac{1}{2}) i \pi \Rightarrow \{ v \to 1/v, \ \ z \to -z, 
\ \ u \to u\}
\label{shift2}
\eeq
where $n$ is an integer.  On a lattice with an even coordination number $q$, 
it is easily seen that these symmetries imply that the magnetisation and
susceptibility are functions only of $u$.  
Complex-temperature curves where the free energy is non-analytic were first
determined for the square lattice by Fisher \cite{fisher1}.  
 For the triangular lattice, the natural boundaries of the free energy, which
also serve as the boundaries of the complex-temperature phases, are 
determined in a similar way as the locus 
of points where the argument of the logarithm vanishes in the (reduced) free 
energy\footnote{
The free energy is trivially infinite at $K=\infty$; since this is an isolated
point and hence does not form part of a natural boundary separating phases, it
will not be important here.}
\beq
f = \ln 2 + \frac{1}{2}\int_{-\pi}^{\pi}\int_{-\pi}^{\pi} \frac{d\theta_1
d\theta_2}{(2\pi)^2} \ln \Bigl [\cosh^3(2K) + \sinh^3(2K) - \sinh(2K)
P(\theta_1,\theta_2) \Bigr ] 
\label{ftri}
\eeq
where 
\beq
P(\theta_1,\theta_2) = \cos \theta_1 + \cos \theta_2 + 
\cos(\theta_1 + \theta_2)
\label{ptheta}
\eeq
Expressed as a function of $u$, this is
\beq
f = 3K + \frac{1}{2}\int_{-\pi}^{\pi}\int_{-\pi}^{\pi} \frac{d\theta_1
d\theta_2}{(2\pi)^2} \ln \Bigl [(1+3u^2) - 2u(1-u)P(\theta_1,\theta_2) \Bigr ] 
\label{ftriu}
\eeq
The function $P(\theta_1,\theta_2)$ 
ranges from a maximum value of 3 at $\theta_1=\theta_2=0$ to a minumum
value of $-3/2$ at $\theta_1=\theta_2=2\pi/3$.  Hence the above 
curves are the locus of solutions to the equation 
\beq
1+3u^2 -2u(1-u)x = 0 \ \quad for \ \quad -\frac{3}{2} \le x \le 3
\label{ueq}
\eeq
This locus consists of the union of the circle 
\beq
u = -\frac{1}{3} + \frac{2}{3}e^{i\phi} 
\label{ucircle}
\eeq
for $0 \le \phi < 2\pi$ with the semi-infinite line segment 
\beq
-\infty \le u \le -\frac{1}{3}
\label{usegment}
\eeq
as pictured in Fig. 1(a).  We label the endpoint ($e$) of this semi-infinite 
line as $u_e$, given above in eq. (\ref{ue}). 
The intersection of the circle (\ref{ucircle}) and the line segment
(\ref{usegment}) consists of the single point 
\beq
u_s = -1
\label{us}
\eeq
which will also be important in our later discussion. 
Since eq. (\ref{ueq}) has real coefficients, the
solutions are either both real or are complex conjugate pairs, which explains
the reflection symmetry of Fig. 1(a) about the $Im(u)=0$ axis. For $x=3$, 
eq. (\ref{ueq}) has a double root at the physical critical point 
$u=u_c=1/3$.  As $x$ decreases from 3 to 0, the complex conjugate roots move
from $u=1/3$ along the circle (\ref{ucircle}) to $\pm i/\sqrt{3}$.  As $x$
decreases from 0 to $-1$, the roots move from these points around the
circle and finally join again to form a double root at $u=-1$ for $x=-1$.  The
behaviour is qualitatively different as $x$ decreases from $-1$ to its minimum
value $-3/2$: in this interval, the solutions consist of two real roots.  These
move outward from $u=-1$ along the negative real axis. As $x \to -3/2$, one
root moves rightward to $u=-1/3$ while the other moves leftward to
$u=-\infty$.  

   The curves in Fig. 1(a) divide the complex $u$ plane into two regions which
are complex-temperature extensions of physical phases: (i) the complex FM
phase, which is the extension of the physical FM phase occupying the interval
$0 \le u \le u_c$, and (ii) the complex PM phase, which is the extension of the
physical PM phase occupying the interval $u_c < u \le 1$.
As is evident from Fig. 1(a), there is a section of the line segment
(\ref{usegment}) protruding into the interior of the complex FM phase.  All
directions of approach to the endpoint $u_e$ of this line segment are from
within the complex FM phase, except for the approach precisely from the left, 
along the line segment itself.  Note that the outer part of the line segment
extending to $u=-\infty$ prevents one from making a circuit around the origin 
staying within either the complex FM or PM phase.  

     This suffices for our analysis of $\bar\chi$, since
it is only a function of $u$. However, especially for our study of
complex-temperature singularities of the Ising model on the 
honeycomb lattice in a companion paper, it is also useful to exhibit the 
corresponding curves in the variable $z$ and $v$. The former are directly
obtained from eqs. (\ref{ucircle})-(\ref{usegment}) and are shown in
 Fig. 1(b). One observes that, whereas the complex $u$-plane consists of two
phases, the complex $z$-plane is divided into three phases; in addition to the
complex-temperature extensions of the PM and FM phase, there is also a third,
denoted O (for ``other'') in Fig. 1(b) lying outside of the boundary 
of the complex FM phase and to the left of the semi-infinite vertical lines
running upward from $z_{e+}=i/\sqrt{3}$ to $i \infty$ and downward from
$z_{e-}=-i/\sqrt{3}$ to $-i\infty$. 
The physical phases are, of course, the (i) FM
phase, in the interval $0 \le z \le z_c = 1/\sqrt{3}$ and (ii) the PM phase, in
the interval $z_c \le z \le \infty$. The O phase has no overlap with any
physical phase.  The same type of phenomenon was true of the
complex-temperature phases for the square lattice (see Fig. 1(b) in our
previous paper I).  

   In order to obtain the analogous boundary curves in the $v$ plane, one 
may either use the bilinear conformal mapping (\ref{bilinearmap}) on the 
curves in Fig. 1(b) or express the free energy in terms of $v$ and solve for 
the points where the argument of the logarithm in the integrand vanishes.  
One finds that, in addition to the isolated point $v=1$ ($K=\infty$), $f$ is 
non-analytic along the locus of points satisfying the equation
\beq
1-2v+6v^2-2v^3+v^4 - 2v(1-v)^2 x = 0  \ \qquad -3/2 \le x \le 3
\label{veq}
\eeq
where, as above, $x=P(\theta_1,\theta_2)$. 
Since this is a quartic equation with real coefficients, the solution for each
$x$ consists generically of two pairs of complex-conjugate roots (which may
include real double roots).  Furthermore, under the mapping $v \to 1/v$,
the left-hand side of eq. (\ref{veq}) transforms into itself, up to an overall
factor of $v^{-4}$.  This implies that the solutions of this equation occur in
reciprocal pairs.  We can describe the
solutions as follows. For $x=3$, there are two solutions: the physical
critical point $v_c=3-2\sqrt{2}$, and its inverse, $1/v_c=3+2\sqrt{2}$. As 
$x$ decreases through the range from 3 to $-1$, the double root at $v=1/v_c$
splits into a conjugate pair, the members of which move along the outer curves
in Fig. 1(c) from $1/v_c$ over to $i$ and $-i$, respectively, while the 
double root at $v_c$ splits into another complex conjugate pair, the
members of which move along the inner curve in this figure, to $-i$ and $i$.
 For $x=-1$, two roots coalesce at $i$ and the other two at $-i$ to form two
double roots.  As $x$ decreases from $-1$ to its minimum, $-3/2$, these
double roots split into four roots, which move outward from $\pm i$ along the
unit circle $|v|=1$.  At $x=-3/2$, the rightward-moving roots reach 
the values $\exp(\pm i\pi/3)$, while the leftward-moving roots coalesce into a
single double root at $v=-1$.  The physical phases are (i) PM, for $0 \le v \le
v_c$, and (ii) FM, for $v_c \le v \le 1$.  The regions marked PM and FM in
 Fig. 1(c) are the extensions of these to complex-temperature.  The outermost 
region in the figure, marked O is a phase which has no overlap 
with any physical phase.  The point $u=u_s=-1$, or equivalently, the points
$z=\pm i$, $v = \mp i$ are singular points of the curves, in the terminology 
of algebraic geometry\cite{alg}; specifically, they are multiple points of 
index 2.  The singular points of the natural boundaries for the Ising model on
the square lattice also occur at these points and are also of index 2.  
However, the topology of the complex phases is different for the triangular and
square lattices, reflecting the fact that the former is a tight-packed 
lattice without an AFM phase, while the latter is a loose-packed lattice. 

  As noted in paper I, using the general fact that the high-temperature and 
(for discrete spin models such as the Ising model) the 
low-temperature expansions have finite radii of convergence, we can use 
standard analytic continuation arguments to establish that not just the free 
energy and its derivatives, but also the magnetisation and susceptibility, 
are analytic functions within each of the complex-temperature extensions of the
physical phases.  This
defines these functions as analytic functions of the respective complex
variable ($K$, $v$, $z$, $u$, or others obtained from these).  We refer the
reader to paper I for other relevant background, definitions, and notation. 

\section{Analysis of Low-Temperature Series}
\label{analysis}

\subsection{Generalities}
\label{general_low}

     The low-temperature series expansion for $\bar\chi$ on the triangular
lattice is given by 
\beq
\bar\chi = 4u^3(1 + \sum_{n=1}^{\infty} c_n u^n)
\label{chiut}
\eeq
This expansion
has a finite radius of convergence and, by analytic continuation from the
physical low-temperature interval $0 \le u < u_c$, applies throughout 
the complex extension of the FM phase.  Since the $u^3$ factor is known 
exactly, it is convenient to study the remaining factor $\bar\chi_r = 
u^{-3}\bar\chi$. For the triangular lattice, following earlier work
\cite{sykes1}, the expansion coefficients $c_{n,t}$ were calculated by the 
King's College group to order $n=13$ (i.e., $\bar\chi$ to $O(u^{16})$) in 
1971 \cite{tlow1}, and to order $n=18$ (i.e., $\bar\chi$ to $O(u^{21})$) in 
1975 \cite{tlow2}.  As discussed above, it was noticed quite early that the 
low-temperature series does not give very good results for the position of 
the physical critical point or for its exponent (as this point is approached 
from within the FM phase), $\gamma'$.  The cause for this was recognized to 
be the existence of the unphysical, complex-temperature singularity at $u_e$, 
the same distance from the origin as the physical singularity at $u_c$ 
\cite{g69,dg}.  We see that this is precisely the endpoint
of the semi-infinite line segment discussed in section \ref{general}, which
protrudes into the interior of the complex FM phase. Since at that time a 
primary goal was to study the physical critical point and, e.g., test the 
lattice universality of critical exponents and various scaling relations, the 
complex-temperature singularity was understandably regarded as a nuisance, and 
methods were devised to transform the series expansion variable so as to 
map it farther from the origin than the physical critical point.  We have noted
above that Guttmann, continuing his earlier work on complex-temperature
singularities of the Ising model on various lattices \cite{g69,dg}, carried 
out a study of this singularity at $u=-1/3$ on the triangular lattice and
determined the corresponding exponent and critical amplitude, using ratio and
dlog Pad\'{e} methods. 

    Here we shall re-analyse this singularity using, in addition to dlog
Pad\'{e} approximants the powerful method of differential approximants.  (For
reviews of these methods, see Refs.\cite{gg}, \cite{pade}, and \cite{tonyg}.)
Given a knowledge of the locus of points where the free energy is non-analytic 
and where the magnetisation vanishes continuously, we have also been motivated
to explore the behaviour of $\bar\chi$ at the point $u_s=-1$. 
Of course, a first thing to check is whether the low-temperature series for 
$\bar\chi$ on the triangular lattice has been calculated to higher order 
since the King's College work in 1975; we have found
that apparently it has not \cite{gaunt,guttmann}.
 As one approaches a generic complex singular point denoted 
$sing$ from within the complex-temperature extension of the FM phase,  
$\bar\chi$ is assumed to have the leading singularity 
\beq
\bar\chi(u) \sim A_{sing}'(1-u/u_{sing})^{-\gamma_{sing}'}
\Bigl ( 1+a_1(1-u/u_{sing}) + ... \Bigr )
\label{singformu}
\eeq
where $A_{sing}'$ and $\gamma_{sing}'$ denote, respectively, the critical 
amplitude and
the corresponding critical exponent, and the $...$ represent analytic confluent
corrections.  One may observe that we have not included non-analytic confluent
corrections to the scaling form in eq. (\ref{singform}).  The reason is that,
although such terms are generally present at critical points in statistical
mechanical models, previous studies have indicated that they are very
weak or absent for the usual critical point of the 2D Ising model
\cite{conflu1,conflu2}.  
The dlog Pad\'{e} and differential approximants then directly give $u_{sing}$ 
and $\gamma_{sing}'$.  As noted above, since the prefactor $u^3$ is known 
and is analytic, we actually study $\bar\chi_r$.  

\subsection{Exponent $\gamma_e'$ at $u=-1/3$, Revisited}
\label{gammaesection}

    We first re-examine the singularity at $u=-1/3$ previously studied in 
detail by Guttmann \cite{g75}.  This singularity is studied by an analysis of
the low-temperature series expansion for $\chi$, since the approach to this
point from all directions in the complex $u$ plane except directly from the
left along the singular line segment (\ref{usegment}) are from with the
(complex-temperature extension of the) FM phase. 
 For exploratory purposes, we first carried out a 
simple dlog Pad\'{e} study of the series for $\bar\chi_r$ in the variable 
$u$.  While this yielded a reasonably stable location for the singularity, 
it did not yield very stable or accurate values for the exponent.  An obvious 
reason for this is the presence of the physical critical point the same 
distance away from the origin.  Accordingly, we transformed the series to one 
in the variable 
\beq
\tilde u=\frac{u}{1-3u}
\label{up}
\eeq
which has the effect of mapping the physical critical point to 
$\tilde u=\infty$ and the complex-temperature singularity at $u_e$ to 
$\tilde u_e=-1/6=-0.16666...$. A dlog Pad\'{e} study of this series gives 
improved results for the location and exponent of the singularity.  We 
obtained by far the most precise results using differential approximants. 
As is well known, this method is more powerful than dlog Pad\'{e}
approximants since it is able to represent an analytic background term as well
as a singular term in a given function.  In this method, the function 
$f = \bar\chi_r(\zeta)$ being approximated satisfies a linear ordinary 
differential equation (ODE) of $K'th$ order, ${\cal L}_{{\bf M},L}f_K(\zeta) =
\sum_{j=0}^{K}Q_{j}(\zeta)D^j f_K(\zeta) = R(\zeta)$, where
$Q_j(\zeta) = \sum_{\ell=0}^{M_j} Q_{j,\ell}\zeta^{\ell}$ and
$R(\zeta)=\sum_{\ell=0}^L R_\ell \zeta^\ell$ (and $\zeta$ denotes a generic
complex variable).  We shall use the implementation of the
method in which $D \equiv \zeta d/d\zeta$.  The solution to 
this ODE, with the initial condition $f(0)=1$, is
the resultant approximant, labelled as $[L/M_0;...;M_K]$.  The general solution
of the ODE has the form $f_j(\zeta) \sim A_j(\zeta)|\zeta-\zeta_j|^{-p_j} +
B(\zeta)$ for $\zeta \to \zeta_j$.  The singular points $\zeta_j$ are
determined as the zeroes of $Q_K(\zeta)$ and are regular singular points of
the ODE, and the exponents are given by $-p_j = K-1-Q_{K-1}(\zeta_j)/(\zeta_j
Q_K'(\zeta_j))$.  
Given the above-mentioned evidence that for the 2D Ising model, 
non-analytic confluent singularities are weak or absent for the physical 
critical point, which motivates the form (\ref{singform}), $K=1$ differential
approximants should be sufficient for our purposes here.  (Of course, $K=2$
differential approximants could be used in a further study to investigate
confluent singularities.)  We also choose to use unbiased differential
approximants since we wish to confirm that the location of the singularity is,
indeed, at $u=-1/3$ as expected.  The results of this analysis are given in
Table 1.  As expected, we obtain higher precision results using differential 
approximants on the series for $\bar\chi_r$ in the transformed variable $\tilde
u$. These are given in Table 2. 

\begin{table}
\begin{center}
\begin{tabular}{|c|c|c|c|} \hline \hline  & & & \\
$[L/M_0;M_1]$ & $u_{sing}$ & $|u_{sing}-u_e|/|u_e|$ & $\gamma_e'$ \\
 & & & \\
\hline \hline

$[0/7;7]$ & $-0.3332767$ & $1.7 \times 10^{-4} $ & $   1.2426$ \\ \hline
$[0/7;8]$ & $-0.3333619$ & $0.86 \times 10^{-4} $ & $  1.2522$ \\ \hline
$[0/7;9]$ & $-0.3333237$ & $2.9 \times 10^{-5} $ & $   1.2476$ \\ \hline
$[0/8;7]$ & $-0.3333691$ & $1.1 \times 10^{-4} $ & $   1.2530$ \\ \hline
$[0/8;8]$ & $-0.3333447$ & $3.4 \times 10^{-5} $ & $   1.2502$ \\ \hline
$[0/9;7]$ & $-0.3333107$ & $0.68 \times 10^{-4} $ & $  1.2460$ \\ \hline
$[1/6;6]$ & $-0.3334340$ & $3.0 \times 10^{-4} $ & $   1.2594$ \\ \hline
$[1/6;7]$ & $-0.3332574$ & $2.3 \times 10^{-4} $ & $   1.2407$ \\ \hline
$[1/6;8]$ & $-0.3333707$ & $1.1 \times 10^{-4} $ & $   1.2534$ \\ \hline
$[1/7;6]$ & $-0.3332608$ & $2.2 \times 10^{-4} $ & $   1.2410$ \\ \hline
$[1/7;7]$ & $-0.3333296$ & $1.1 \times 10^{-5} $ & $   1.2478$ \\ \hline
$[1/7;8]$ & $-0.3334177$ & $2.5 \times 10^{-4} $ & $   1.2602$ \\ \hline
$[1/8;6]$ & $-0.3333782$ & $1.3 \times 10^{-4} $ & $   1.2542$ \\ \hline
$[2/6;6]$ & $-0.3334040$ & $2.1 \times 10^{-4} $ & $   1.2581$ \\ \hline
$[2/6;7]$ & $-0.3333311$ & $0.68 \times 10^{-5} $ & $  1.2494$ \\ \hline
$[2/6;8]$ & $-0.3333613$ & $0.84 \times 10^{-4} $ & $  1.2524$ \\ \hline
$[2/7;6]$ & $-0.3333310$ & $0.69 \times 10^{-5} $ & $  1.2494$ \\ \hline
$[2/7;7]$ & $-0.3333301$ & $0.96 \times 10^{-5} $ & $  1.2484$ \\ \hline
$[2/8;6]$ & $-0.3333618$ & $0.85 \times 10^{-4} $ & $  1.2525$ \\ \hline
$[3/4;6]$ & $-0.3333208$ & $3.8 \times 10^{-5} $ & $   1.2282$ \\ \hline
$[3/6;6]$ & $-0.3332486$ & $2.5 \times 10^{-4} $ & $   1.2374$ \\ \hline
$[4/6;4]$ & $-0.3333540$ & $0.62 \times 10^{-4} $ & $  1.2553$ \\ \hline
$[5/5;6]$ & $-0.3332986$ & $1.0 \times 10^{-4} $ & $   1.2423$ \\ \hline
$[6/4;6]$ & $-0.3333335$ & $0.47 \times 10^{-6} $ & $  1.2463$ \\ \hline
$[6/5;5]$ & $-0.3332566$ & $2.3 \times 10^{-4} $ & $   1.2375$ \\ \hline
$[6/6;4]$ & $-0.3334199$ & $2.6 \times 10^{-4} $ & $   1.2624$ \\ \hline
$[7/4;4]$ & $-0.3333719$ & $1.2 \times 10^{-4} $ & $   1.2570$ \\ \hline
$[8/4;3]$ & $-0.3333846$ & $1.5 \times 10^{-4} $ & $   1.2596$ \\ \hline
$[8/4;4]$ & $-0.3333423$ & $2.7 \times 10^{-5} $ & $   1.2511$ \\ \hline

\end{tabular}
\end{center}
\caption{Values of $u_{sing}$ and $\gamma_e$ from differential
approximants to low-temperature series for $\bar\chi_r(u)$. See text for
definition of $[L/M_0,M_1]$ approximant.  We only display entries which 
satisfy the accuracy criterion $|u_{sing}-u_e|/|u_e| \le 3 \times 10^{-4}$
where $u_e=-1/3$.}
\label{table1}
\end{table}

\begin{table}
\begin{center}
\begin{tabular}{|c|c|c|c|} \hline \hline  & & & \\
$[L/M_0;M_1]$ & $\tilde u_{sing}$ & 
$|\tilde u_{sing}-\tilde u_e|/|\tilde u_e|$ & $\gamma_e'$ \\
 & & & \\
\hline \hline

$[0/6;7]$ & $-0.1666760$ & $0.56 \times 10^{-4}$   & $1.2516$ \\ \hline
$[0/6;8]$ & $-0.1666666$ & $1.0  \times 10^{-6}$   & $1.2481$ \\ \hline
$[0/7;6]$ & $-0.1666760$ & $0.56 \times 10^{-4}$   & $1.2517$ \\ \hline
$[0/7;7]$ & $-0.1666643$ & $1.4 \times 10^{-5}$    & $1.2471$ \\ \hline
$[0/7;8]$ & $-0.1666621$ & $2.7 \times 10^{-5}$    & $1.2462$ \\ \hline
$[0/7;9]$ & $-0.1666628$ & $2.3 \times 10^{-5}$    & $1.2466$ \\ \hline
$[0/8;6]$ & $-0.1666666$ & $4.9 \times 10^{-7}$    & $1.2481$ \\ \hline
$[0/8;7]$ & $-0.1666621$ & $2.7 \times 10^{-5}$    & $1.2462$ \\ \hline
$[0/8;8]$ & $-0.1666629$ & $2.3 \times 10^{-5}$    & $1.2466$ \\ \hline
$[0/9;7]$ & $-0.1666628$ & $2.3 \times 10^{-5}$    & $1.2466$ \\ \hline
$[1/6;6]$ & $-0.1666543$ & $0.74 \times 10^{-4}$   & $1.2424$ \\ \hline
$[1/6;7]$ & $-0.1666587$ & $0.48 \times 10^{-4}$   & $1.2443$ \\ \hline
$[1/6;8]$ & $-0.1666584$ & $0.50 \times 10^{-4}$   & $1.2441$ \\ \hline
$[1/7;6]$ & $-0.1666587$ & $4.8 \times 10^{-5}$    & $1.2443$ \\ \hline
$[1/7;7]$ & $-0.1666582$ & $5.1 \times 10^{-5}$    & $1.2440$ \\ \hline
$[1/7;8]$ & $-0.1666749$ & $4.9 \times 10^{-5}$    & $1.2535$ \\ \hline
$[1/8;6]$ & $-0.1666583$ & $5.0 \times 10^{-5}$    & $1.2441$ \\ \hline
$[1/8;7]$ & $-0.1666748$ & $4.9 \times 10^{-5}$    & $1.2535$ \\ \hline
$[2/5;7]$ & $-0.1666817$ & $0.90 \times 10^{-4}$   & $1.2564$ \\ \hline
$[2/6;6]$ & $-0.1666742$ & $0.45 \times 10^{-4}$   & $1.2524$ \\ \hline
$[2/6;7]$ & $-0.1666577$ & $0.54 \times 10^{-4}$   & $1.2438$ \\ \hline
$[2/6;8]$ & $-0.1666630$ & $2.2 \times 10^{-5}$    & $1.2467$ \\ \hline
$[2/7;5]$ & $-0.1666817$ & $0.90 \times 10^{-4}$   & $1.2565$ \\ \hline
$[2/7;6]$ & $-0.1666577$ & $0.54 \times 10^{-4}$   & $1.2438$ \\ \hline
$[2/7;7]$ & $-0.1666619$ & $2.8 \times 10^{-5}$    & $1.2461$ \\ \hline
$[2/8;6]$ & $-0.1666631$ & $2.2 \times 10^{-5}$    & $1.2467$ \\ \hline

\end{tabular}
\end{center}
\caption{Values of $\tilde u_{sing}$ and $\gamma_e'$ from differential
approximants to low-temperature series for $\bar\chi_r(\tilde u)$. See text for
definition of $[L/M_0,M_1]$ approximant.  We only display entries which 
satisfy the accuracy criterion $|u_{sing}-u_e|/|u_e| \le 1 \times 10^{-4}$.}
\label{table2}
\end{table}

\begin{table}
\begin{center}
\begin{tabular}{|c|c|c|c|} \hline \hline  & & & \\

$[3/5;6]$ & $-0.1666780$ & $0.68 \times 10^{-4}$   & $1.2545$ \\ \hline
$[3/5;7]$ & $-0.1666593$ & $0.44 \times 10^{-4}$   & $1.2446$ \\ \hline
$[3/6;5]$ & $-0.1666780$ & $0.68 \times 10^{-4}$   & $1.2545$ \\ \hline
$[3/6;6]$ & $-0.1666564$ & $0.61 \times 10^{-4}$   & $1.2431$ \\ \hline
$[3/6;7]$ & $-0.1666615$ & $3.1 \times 10^{-5}$    & $1.2458$ \\ \hline
$[3/7;5]$ & $-0.1666593$ & $0.44 \times 10^{-4}$   & $1.2446$ \\ \hline
$[3/7;6]$ & $-0.1666615$ & $3.1 \times 10^{-5}$    & $1.2458$ \\ \hline
$[4/5;5]$ & $-0.1666748$ & $0.49 \times 10^{-4}$   & $1.2528$ \\ \hline
$[4/5;6]$ & $-0.1666575$ & $0.55 \times 10^{-4}$   & $1.2436$ \\ \hline
$[4/5;7]$ & $-0.1666632$ & $2.1 \times 10^{-5}$    & $1.2468$ \\ \hline
$[4/6;5]$ & $-0.1666575$ & $0.55 \times 10^{-4}$   & $1.2436$ \\ \hline
$[4/6;6]$ & $-0.1666593$ & $0.44 \times 10^{-4}$   & $1.2444$ \\ \hline
$[4/7;5]$ & $-0.1666633$ & $2.0 \times 10^{-5}$    & $1.2469$ \\ \hline
$[5/5;5]$ & $-0.1666577$ & $0.54 \times 10^{-4}$   & $1.2437$ \\ \hline
$[5/5;6]$ & $-0.1666634$ & $1.9 \times 10^{-5}$    & $1.2470$ \\ \hline
$[5/6;4]$ & $-0.1666809$ & $0.85 \times 10^{-4}$   & $1.2573$ \\ \hline
$[5/6;5]$ & $-0.1666637$ & $1.8 \times 10^{-5}$    & $1.2472$ \\ \hline
$[6/4;6]$ & $-0.1666618$ & $2.9 \times 10^{-5}$    & $1.2458$ \\ \hline
$[6/5;3]$ & $-0.1666838$ & $1.0 \times 10^{-4}$    & $1.2594$ \\ \hline
$[6/5;5]$ & $-0.1666630$ & $2.2 \times 10^{-5}$    & $1.2467$ \\ \hline
$[6/6;4]$ & $-0.1666672$ & $3.2 \times 10^{-6}$    & $1.2496$ \\ \hline
$[7/4;5]$ & $-0.1666595$ & $0.43 \times 10^{-4}$   & $1.2444$ \\ \hline
$[7/5;4]$ & $-0.1666691$ & $1.5 \times 10^{-5}$    & $1.2507$ \\ \hline
$[8/5;3]$ & $-0.1666811$ & $0.87 \times 10^{-4}$   & $1.2578$ \\ \hline

\end{tabular}
\end{center}
\caption{continuation of Table 2.}
\label{table4cont}
\end{table}

   The results of this study agree with the old inference of a singularity in
$\bar\chi$ at $u=u_e=-1/3$ \cite{dg} (i.e., $\tilde u=-1/6$).  We have found
that the differential approximants yield a considerably more precise
determination of the exponent than the dlog Pad\'{e} approximants, and our
results are in excellent agreement with Guttmann's earlier conclusion using
ratio and dlog Pad\'{e} methods that $\gamma_e'=5/4$ \cite{g75}.  
Specifically, we find 
\beq
\gamma_e' = 1.249 \pm 0.005
\label{gammaenum}
\eeq
As will be discussed in our companion paper on the honeycomb lattice
\cite{chihc}, the result (\ref{gammaenum}) is closely related to two findings 
which we report there, viz., that for the Ising model on the honeycomb lattice,
the (uniform) susceptibility $\bar\chi$ and the staggered susceptibility
$\bar\chi^{(a)}$ have divergent singularities at the point $z=z_{\ell}=-1$ 
with respective exponents which we infer to have the values 
$\gamma_{\ell}'=\gamma_{\ell,a}'=2\gamma_{e}'=5/2$.

\subsection{Critical Amplitude at $u=-1/3$ Singularity}
\label{ae}

    Here we have used a method complementary to that employed earlier by 
Guttmann to extract the critical amplitude $A_e'$ \cite{g75}.  
Specifically, we 
compute the series for $(\bar\chi_r)^{1/\gamma_e'}$.  Since the exact function 
$(\bar\chi_r)^{1/\gamma_e'}$ has a simple pole at $u_e$, one performs a
Pad\'{e} analysis on the series itself instead of its logarithmic derivative.
The residue at this pole is 
\beq
R_e = -u_e(A'_{r,e})^{1/\gamma_e'}
\label{re}
\eeq
where $A'_{r,e}$ denotes the critical amplitude for $\bar\chi_r$.  It follows
that 
\beq
A_e' = 4u_e^3 A_{r,e}'= -4(3)^{-7/4}R_{e}^{5/4}
\label{aep}
\eeq
Using the inferred value $\gamma_e'=5/4$ to compute the series, extracting the
residue $R_e$, and then using eq. (\ref{aep}), we obtain the critical
amplitude.  We have also done the same analysis on the transformed series in
the variable $\tilde u = u/(1-3u)$ for which the physical critical point is
mapped to infinity, and $u_e$ to $\tilde u_e=-1/6$.  Equating the singular 
forms
\beqs
\bar\chi(u) & \sim & A_e'(1-u/u_e)^{-5/4} \nonumber \\
& = & \bar\chi(\tilde u) \sim \tilde A_e'(1-\tilde u/\tilde u_e)^{-5/4}
\label{singeq}
\eeqs
we have 
\beq
A_e' = 2^{5/4}\tilde A_e'
\label{aeaetilde}
\eeq
In our analysis of the Pad\'{e} results, we plot the individual determinations
of the critical amplitude from each $[L/M]$ approximant as a function of the
accuracy with which that approximant locates the singular point, as measured by
the relative deviation $|u_{sing}-u_e|/|u_e|$ and 
$|\tilde u_{sing} - \tilde u_e|/|\tilde u_e|$.  We obtain the final estimate of
the critical amplitude by extrapolating these results to zero deviations. 
Combining our results from both analyses, we obtain the result 
\beq
A_e' = -0.05766 \pm 0.00015
\label{aenum}
\eeq
This agrees with the result by Guttmann \cite{g75}, cited in eq. (\ref{aeg}) 
and has a somewhat smaller estimated uncertainty. 

   It is of interest to compare this with the critical amplitude at the
physical critical point, as approached from within the FM phase.  For
reference, we note that many authors express the singularity in terms of $T$ 
rather than $u$, viz.,
\beqs
\bar\chi_{sing} & \sim & A_{c,T}'(1-T/T_c)^{-\gamma'} \nonumber \\
                & \sim & A_c'(1-u/u_c)^{-\gamma'} 
\label{singformt}
\eeqs
with $\gamma'=7/4$.  The critical amplitudes are related according to 
\beq
A'_c = (4K_c)^{7/4}A'_{c,T}
\label{amprel}
\eeq
An early analysis by Essam and Fisher of a low-temperature series for 
$\bar\chi$ up to $O(u^{10})$ for $\bar\chi$ yielded the result \cite{ef} 
$A_{c,T}'=0.0248 \pm 0.0006$, 
or equivalently, $A_c'=0.0292 \pm 0.0007$.  A more accurate determination 
was made by Guttmann \cite{gca}, based on the series for 
$\bar\chi$ calculated to order $O(u^{21})$ \cite{tlow2}, with the result 
$A_{c,T}'=0.0246 \pm 0.0002$, i.e., $A_c=0.0290 \pm 0.0002$.  In the same
paper, using a generalised law of corresponding states, as formulated in
Ref. \cite{bgj} (further discussed in Ref. \cite{law}), 
Guttmann obtained a high-precision analytic determination of
the critical amplitudes for the Ising susceptibility for the triangular,
honeycomb, and kagom\'{e} lattices (analogous to that obtained in
Ref. \cite{conflu1} for the square lattice).  For the triangular lattice, he
calculated 
\beq
A_{c,T}' = 0.0245189020 \, \qquad i.e., \ A_c'=0.028905388
\label{ac}
\eeq
Using this in conjunction with eq. (\ref{aenum}), we find 
\beq
\frac{A_e'}{A_c'} = -(1.995 \pm 0.005)
\label{amprat}
\eeq
where the uncertainty arises completely from the uncertainty in $A_e'$.  One
may observe that the ratio (\ref{amprat}) is slightly less than, but close 
to, the simple relation $A_e'/A_c' = -2$.  We should note, however, that we 
do not know of a specific analytic basis for such a simple amplitude relation. 
 
\subsection{On the Singularity at $u=-1$} 

   Although the complex-temperature phase diagram of the Ising model on the
triangular lattice is different from that on the square lattice, the point 
$u=u_s=-1$ is somewhat similar for both of these lattices, in the sense that
at this point on the two respective lattices, (i) the spontaneous 
magnetisation and the inverse correlation length vanish continuously as one 
approaches this point from within the complex FM phase (as will be discussed
further below).  The differences include the fact that (ii) on the square
lattice, the complex PM, FM, and AFM phases are all contiguous at $u=-1$,
whereas on the triangular lattice, only the complex FM and PM phases are
contiguous (there is no AFM phase); and (iii) on the square lattice, the 
point $u=-1$ can be reached by moving in a straight line outward along the 
negative real axis from the origin in the $u$-plane, whereas on the 
triangular lattice, it cannot be; 
one first encounters the endpoint of the singular semi-infinite line 
segment (\ref{usegment}) at $u=-1/3$.  It is thus naturally of interest to
investigate the behaviour of the susceptibility at $u=-1$. 

   For this purpose, we have carried out an analysis of the low-temperature 
series for $\bar\chi$ on the triangular lattice, using dlog Pad\'{e} and 
differential approximant methods.  However, we have found that the 
series, at least at the order to which it has been calculated, cannot probe
this point very sensitively.  We believe that the reason for this is the 
property (iii) listed above, viz., the fact that the series is dominated by the
singularity at $u=-1/3$, which is not only closer to the origin but also 
directly in
front of the point $u=-1$ as approached from the origin.  This is just as true
of the series transformed to the variable $\tilde u$ as it is of the series in
$u$ since the image of the point $u=-1/3$ is at $\tilde u=-1/6$, directly in
front of the image of the point $u=-1$, at $\tilde u=-1/4$, as approached from
the origin in the $\tilde u$ plane.  Hence, one does not expect the series
transformed to the $\tilde u$ variable to be more sensitive to the $u=-1$
singularity, and, indeed, it is not.  We have analysed the series with many
transformations of expansion variables which are designed to map both the
$u=-1/3$ singularity and the physical critical point at $u=1/3$ farther away
from the origin than the $u=-1$ singularity.  Examples include $u_{trans.} =
\sinh(u)/(1-9u^2)$ and $u_{trans.} = u \exp(a (1+u)^p - a)$ for various $a > 0$
and positive even integer $p$.  However, even in the transformed variables, 
the series yield inconclusive results for $\gamma_s'$.  
We can say that if the scaling 
relation $\alpha'+2\beta+\gamma'=2$ holds at $u=-1$ for the Ising model on the
triangular lattice, as it does on the square lattice, then, given the exact
result that $\beta_s=3/8$ (see eq. (\ref{mtri}) below) and our finding that
$\alpha_s'= 0$ from the exact free energy (see eq. (\ref{alphas}) below), it
would follow that $\gamma_s'=5/4$, which would be equal to the value of the
exponent $\gamma_s'$ for the singularity at $u_e$. 

    In passing, we note that the analogue of the transformation to an elliptic
modulus variable which we found to be very valuable in the case of the
square lattice is not helpful here, for reasons which can easily be explained. 
The requisite elliptic modulus is clear from the exact solutions for the free
energy \cite{ftri} and spontaneous magnetisation \cite{potts52} of the model
(the latter of which can be written as $M=(1-(k_<)^2)^{1/8}$) and is given by 
\beq
k_< = \frac{4u^{3/2}}{(1+3u)^{1/2}(1-u)^{3/2}}
\label{kl}
\eeq
 First, the image of the singular boundary (\ref{ucircle})-(\ref{usegment})
extends all the way in to the origin in the complex $k_<$-plane.  Specifically,
the semi-infinite line segment (\ref{usegment}) is mapped via a 2-valued 
homomorphism to the segment $-1 \le (k_<) \le 0$.  This is quite different 
from the situation for the square lattice, where the mapping from $u$ to the 
elliptic modulus variable for that lattice, viz., $k_< = 4u/(1-u)^2$, mapped 
the boundary of the complex FM phase to the unit circle $|k_<|=1$, so that 
no portion of this boundary 
impinged upon the origin in the complex $(k_<)$-plane.  A second important
difference is that whereas for the 
square lattice, $|k_<|$ is less than unity in 
the complex FM phase, for the triangular lattice, $|k_<|$ is not only not less
than unity in this phase; it actually diverges as one approaches the 
endpoint $u_e$ of the singular line segment protruding into this phase. To see
this, let 
\beq
u = -\frac{1}{3} + \frac{1}{3}\epsilon e^{i \phi}
\label{uform}
\eeq
where $\epsilon$ is real and positive. Then, 
\beq
k_< \to -\frac{i}{2(\epsilon e^{i\phi})^{1/2}} \ \quad \ as \quad 
\epsilon \to 0
\label{klim}
\eeq
Taking the branch cut for the fractional powers in (\ref{kl}) to lie from 
$u=u_e$ to $u= -\infty$, then for the approach to $u=-1/3$ from the origin
along the negative real axis, which corresponds to $\phi=0$ in
eqs. (\ref{uform}) and (\ref{klim}), $k_< \sim -(i/2)(\epsilon)^{-1/2} \to 
-i \infty$.  

\section{Behaviour of Magnetisation and Implications}

\subsection{Divergence of $M$ at $u=-1/3$}

  In discussing our results for the singularity in $\bar\chi$ at 
$u=u_e-1/3$, we 
first note that some insight into the location of this singularity can be 
gained from an examination 
of the exact expression for the spontaneous magnetisation of the Ising 
model on the triangular lattice \cite{potts52}:
\beq
M = \biggl (\frac{1+u}{1-u}\biggr )^{3/8}
\biggl(\frac{1-3u}{1+3u}\biggr)^{1/8}
\label{mtri}
\eeq
which applies within the physical FM phase and, by analytic continuation,
throughout the complex-temperature extension of this phase as shown in
 Fig. 1(a). (Of course, outside of this complex-extended FM phase, $M$ 
vanishes identically.)  $M$ diverges\footnote{
 For physical temperatures, the unit-normalized magnetisation is bounded to 
lie between 0 and 1.  However, this bound does not apply for complex 
temperatures, and, indeed, nothing prevents $M$ from having divergent 
singularities for complex temperatures.} 
at the point $u_e$, with exponent 
\beq
\beta_e = -\frac{1}{8}
\label{betae}
\eeq
Thus it is not surprising that at a point such as $u_e$ where there is a 
singularity in $M$, there is also a related singularity in $\bar\chi$.  

\subsection{Theorem on $M \to \infty \Rightarrow \chi \to \infty$}

Such an exotic phenomenon as a divergent spontaneous magnetisation has received
very little attention in the literature.  We begin by stating and proving a 
theorem which deals with an important effect of such a divergence.  First, we
prove a lemma concerning 2-spin correlation functions:
\vspace{2mm}
\begin{flushleft}
Lemma 1
\end{flushleft}
Assume that a given statistical mechanical model has a phase with ferromagnetic
long-range order.  In this phase, and in its extension to complex
temperature, the 2-spin correlation function can always be written in the form 
\beq
<\sigma_n \sigma_{n'}> = M^2 c(n,n')
\label{mprefactor}
\eeq
where $M$ is the spontaneous magnetisation and $c(n,n')$ contains all of the
dependence on the lattice sites $n$ and $n'$.  
\vspace{2mm}
\begin{flushleft}
Proof
\end{flushleft}
This result follows easily from the fact that one of the equivalent definitions
of $M$ is precisely via the relation
\beq
M^2 = \lim_{|n-n'| \to \infty} <\sigma_n \sigma_{n'}>
\label{msquared}
\eeq
Given the correlation function $<\sigma_n \sigma_{n'}>$, one may thus calculate
$M^2$ from the limit (\ref{msquared}) and divide, thereby obtaining the
function $c(n,n')$ (with the property $\lim_{|n-n'| \to \infty}c(n,n') = 1$).
$\Box$ \newline 
As an immediate corollary, we have 
\vspace{2mm}
\begin{flushleft}
Lemma 2
\end{flushleft}
Assume that a given statistical mechanical model has a phase with ferromagnetic
long-range order.  In this phase, and in its extension to complex
temperature, the connected 2-spin correlation function can always be written
in the form 
\beq
<\sigma_n \sigma_{n'}>_{conn.} = M^2 c(n,n')_{conn.}
\label{mprefactorconn}
\eeq
where $M$ is the spontaneous magnetisation and $c(n,n')_{conn.}$ contains all 
of the dependence on the lattice sites $n$ and $n'$.  
\vspace{2mm}
\begin{flushleft}
Proof
\end{flushleft}
This follows immediately from the definition of the connected 2-spin 
correlation function as $<\sigma_n \sigma_{n'}>_{conn.} \ \equiv \ 
<\sigma_n \sigma_{n'}> - M^2$, which also shows that $c(n,n')_{conn} = 
c(n,n')-1$. $\Box$ \newline
We then proceed to
\vspace{2mm}
\begin{flushleft}
Theorem 1. 
\end{flushleft}
If the magnetisation $M$ diverges as one approaches a given point from within
the (complex-temperature extension of the) ferromagnetic phase, then, in the
same limit, the susceptibility also diverges. 
\vspace{2mm}
\begin{flushleft}
Proof 
\end{flushleft}
The susceptibility $\bar\chi$ is given as the sum over the connected 2-spin 
correlation functions
\beqs
\bar\chi & = & N_s^{-1} \lim_{N_s \to \infty} \sum_{n,n'} 
<\sigma_n \sigma_{n'}>_{conn.} \nonumber \\
         & = & \sum_n <\sigma_0 \sigma_n>_{conn.}
\label{fdthm}
\eeqs
(where the homogeneity of the lattice has been used in the second line).  Using
Lemma 2, we have 
\beq
\bar\chi = M^2 \sum_n c(n,n')_{conn.}
\label{mpref}
\eeq
It follows that, in general, a divergence in $M$ will cause a divergence in
$\bar\chi$.  $\Box$ \newline
Note that this would be true even in the hypothetical case in which $M$ is 
divergent but the correlation length is finite, so that the sum 
$\sum_n c(n,n')_{conn.}$ converges.  In passing, we note that if $M$ diverges
at a given point $u_0$ (or
the corresponding points $z_0$), this will correspond to the infinite set of 
values of $K$ which map to $u_0$ or the set of $z_0$ according to the 
symmetries (\ref{shift1}) and, for even $q$, (\ref{shift2}). 
\vspace{3mm}

We next apply Theorem 1 to our immediate topic of study:
\vspace{2mm}
\begin{flushleft}
Corollary 1.
\end{flushleft}
\vspace{2mm}
 For the Ising model on the triangular lattice, $\bar\chi$ has a divergent
singularity at $u=-1/3$. 
\vspace{2mm}
\begin{flushleft}
Proof
\end{flushleft}
This follows from the fact that, as we know from the exact result,
(\ref{mtri}), $M$ (analytically continued throughout the
complex-temperature extension of the FM phase) diverges at $u=-1/3$ together
with Theorem 1. $\Box$ \newline
Note that unlike the study of the low-temperature series, which is, of 
course, approximate since the series only extends to finite order, this is an 
exact rigorous result.  What the studies by Guttmann \cite{g75} and the present
authors  yielded beyond the result of the theorem was the actual values of 
the exponent $\gamma_e'$ and critical amplitude $A_e'$.  

   Also, observe that we did not explicitly use any 
property of the correlation length to make this conclusion.  Our theorem and
corollary also allow us to infer without a direct calculation that the
correlation length does in fact diverge at $u=-1/3$ (as this point is 
approached from the interior of the complex-temperature extension
of the FM phase, i.e. from all directions except from the left along the 
singular line segment (\ref{usegment})).  We can deduce this
because if the correlation length were finite, then the sum over 2-spin
correlation functions in (\ref{fdthm}) would be finite.  (Since only the
large-distance behaviour is relevant to a possible divergence, one can replace
the sum by an integral, and because of the exponential damping from 
$<\sigma_0 \sigma_r> \sim r^{-p}e^{r/\xi}$, the integral is finite.)  But then
since the only divergence would arise from the prefactor of $M^2$, we would 
have the exponent relation $\gamma_e' = -2\beta_e$.  Since $-2\beta_e=1/4$, 
while the series analyses yield $\gamma_e'=5/4$, the above exponent relation 
does not hold.  This shows then, that the correlation length diverges at 
$u=-1/3$. 

\subsection{Other Singularities of $M$}

    In addition to its divergence at $u=-1/3$, $M$ as given by (\ref{mtri})
vanishes continuously at the complex-temperature point $u=u_s=-1$, with 
exponent 
\beq
\beta_s=\frac{3}{8}
\label{betas}
\eeq
as well as the usual critical point, $u=u_c=1/3$ (with exponent $\beta=1/8$).
With the exception of these
two points, $u_c$ and $u_s$, $M$ vanishes discontinuously elsewhere along 
the outer boundary of the complex-temperature extension of the FM phase.  Note
that despite the $(1-u)^{-3/8}$ factor in the exact expression (\ref{mtri}),
$M$ does not in fact diverge at $u=1$, since this point lies outside the region
where eq. (\ref{mtri}) applies (which, we recall, is the region which can be
reached by analytic continuation from the physical FM phase, i.e. the 
complex-temperature extension of the FM phase). 

\section{Some Remarks on Scaling Relations}

    We have previously shown that in general at complex temperature
singularities, a number of the usual scaling relations applicable for physical
critical points do not hold.  In particular, for the Ising model on the square
lattice we have proved \cite{ms,chisq} that as one approaches the boundary
of the complex-temperature extension of the symmetric, PM phase, the
susceptibility has only finite non-analyticities, except for its divergence at
the physical critical point.  Our proof was based on the fact that in this
limit, the inverse correlation length only vanishes at the physical critical
point.  From our theorem, it follows that at the points $v= \pm i$ (i.e., the
point $u=u_s=-1$), as approached from within the complex PM phase, the exponent
describing the finite non-analyticity in $\bar\chi$ is negative:
$\gamma_s <0$. 
On the other hand, at this point $u_s =-1$, as approached from within the
interior of the complex-temperature extension of the FM phase, $\bar\chi$ does
diverge \cite{egj,chisq}.  As we noted in Ref. \cite{chisq}, this implies,
among other things, that at this singularity, $u=u_s=-1$, $\gamma_s \ne
\gamma_s'$.  This violates the exponent scaling relation (derived from the 
usual renormalization-group method) applicable at physical critical points, 
that $\gamma'=\gamma$.  

    Secondly, we have shown \cite{chisq} by the use of exact results for the 
magnetisation, that universality, in the sense that critical exponents are
independent of lattice type, is, in general, violated at complex-temperature
singularities.  

    Thirdly, we have shown \cite{chisq} that for the square lattice, 
as one approaches the point 
$u=u_s=-1$ from within the complex FM (or AFM) phase, the inverse 
correlation length $\xi_{FM,row}^{-1}$ defined from the 
row (or equivalently, column) connected 2-spin correlation functions 
(analytically continued throughout the complex-temperature extension of the FM
phase) vanishes like $|1+u|^{\nu_s'}$ with $\nu_s'=1$, whereas 
the inverse correlation length $\xi_{FM,d}^{-1}$ defined from the diagonal 
connected 2-spin correlation functions vanishes like $|1+u|^{-\nu_{s,diag.}}$
with $\nu_{s,diag.}=2$ in the same limit.  As noted in paper I, this situation
is unprecedented for physical critical points.  Indeed, more generally, using
the exact 
calculation of the asymptotic form of the 2-spin correlation function 
$<\sigma_{0,0}\sigma_{m,n}>$ \cite{cw} for the square lattice (analytically 
continued throughout the complex FM phase), we have shown that the inverse 
correlation length extracted from 
$<\sigma_{0,0}\sigma_{m,n}>_{conn.}$ as $r=(m^2+n^2)^{1/2} \to \infty$ vanishes
as $u \to u_s$ from within the complex FM phase like $|u+1|^{\nu_s'}$ with 
$\nu_s'=1$ if $\theta = \arctan(m/n)$ does not represent a diagonal of the
lattice, i.e. is not equal to $\pm \pi/2$ or $\pm 3\pi/2$.  
This result means that
for the Ising model on the square lattice, as one approaches $u=-1$ from 
within the complex FM phase, the spin-spin correlations along the diagonal 
become long-ranged faster than those in other directions.  This, in turn,
undermines the naive use of renormalization group methods to derive scaling
relations for exponents since these methods rely on the existence of a single
diverging length scale provided by the correlation length. 

    Nevertheless, one does find \cite{egj,chisq} that for the square lattice,
as one approaches the point $u=-1$ from within the complex FM phase, the
exactly known exponents $\alpha_s'= 0$ (log divergence in specific heat) and
$\beta_s=1/4$, together with the inferred exact result \cite{egj,chisq} 
$\gamma_s'=3/2$, satisfy
the scaling relation $\alpha_s'+2\beta_s+\gamma_s'=2$.  However, as we noted
\cite{chisq}, the corresponding relation for the approach to $u=-1$ from within
the symmetric phase, viz., $\alpha_s+2\beta_s+\gamma_s=2$ is false, since
$\alpha_s=0$ (log divergence) and, as mentioned above, $\gamma_s < 0$.  

    These findings show that universality, scaling, and exponent relations
which were applicable to physical critical points do not, in general, hold for
complex-temperature singular points.  In particular, for the subset of these
complex-temperature singularities where $M$ diverges, there is yet another
unusual implication.  Let us define 
\beq
t \equiv |T-T_c|/T_c
\label{t}
\eeq
As $t \to 0$, the connected 2-spin correlation function approaches the critical
decay 
\beq
<\sigma_0\sigma_r>_{conn.} \to r^{-(d-2+\eta)}
\label{sscritical}
\eeq
In the FM phase, this correlation function can also be written in the form
(\ref{mprefactorconn}), i.e., $<\sigma_n\sigma_{n'}>_{conn.} = M^2 c(n,n')$. 
Then according to standard assumptions of scaling and the renormalization 
group, as $t \to 0$ and $r \to \infty$ with 
\beq
\rho = r/\xi
\label{rho}
\eeq
fixed, where $\xi \sim t^{-\nu}$, the spatially-dependent term $c(n,n')$
becomes only a function of the scaled length, $\rho$, i.e., 
\beq
c(n,n') \sim \rho^{-(d-2+\eta)}
\label{casymp}
\eeq
so that in this limit, 
$r^{-(d-2+\eta)} \sim t^{2\beta}/(rt^{\nu'})^{(d-2+\eta)}$, and hence 
\beq
2\beta=\nu'(d-2+\eta)
\label{scalrelb}
\eeq
which is one of the well-known scaling relations among critical exponents. 
Now recall that this use of the renormalization group requires a divergent
correlation length, so $\nu'$ must be positive.\footnote{As usual, logarithmic
factors could, in principle, be present, in particular, at the upper critical
dimensionality; they would not change the value of the algebraic exponents.}
But if $\beta < 0$, as is the case for a complex-temperature singularity
where $M$ diverges, then eq. (\ref{scalrelb}) implies that $d-2+\eta < 0$, so
that the critical connected 2-spin correlation function, eq. 
(\ref{sscritical}), diverges with the distance $r$ between the spins, rather
than having its usual decay to zero as $r \to \infty$.  Of course, this
behaviour could never happen for physical temperatures, but it is forced on us
at a complex-temperature critical point where (i) $M$ diverges, provided that
(ii) the correlation length also diverges at this point and, (ii) the usual
scaling assumption for the 2-spin correlation function is valid. 

\section{Behaviour of the Specific Heat}

\subsection{General}

   By re-expressing the exact solution \cite{ftri}\footnote{Houtappel's
expressions for the internal energy and specific heat, eqs. (108) and (109),
respectively, in Ref. \cite{ftri}, are incorrect if one uses
the integrals $\epsilon_1(\beta)$ and $\epsilon_2(\beta)$ as he defines them,
with the range of integration from $\phi=0$ to $2\pi$.  If, instead, one takes
the range of integration from $\phi=0$ to $\phi=\pi/2$, so that the integrals
are just the usual elliptic integrals $K(\sqrt{\beta})$ and $E(\sqrt{\beta})$,
then his eqs. (108) and (109) become correct.} for the specific 
heat $C$ for the Ising model on the triangular lattice in the FM phase in 
terms of the variable $u$, we obtain 
\beq
k_B^{-1} K^{-2}C = -\frac{8u}{(1-u)^2} + \frac{2(3u^3+9u^2-7u+3)k_<}
{\pi u^{3/2}(1-u)^{2}}K(k_<) 
- \frac{6(1-u)k_<}{\pi u^{3/2}}E(k_<)
\label{ctri}
\eeq
where $K(k)=\int_{0}^{\pi/2}d\theta [1-k^2\sin^2\theta]^{-1/2}$ and 
$E(k)=\int_{0}^{\pi/2}d\theta [1-k^2\sin^2\theta]^{1/2}$ are the complete 
elliptic integrals of the first and second kinds, respectively, depending 
on the the elliptic modulus $k$, and $k_<$ was given in eq. (\ref{kl}). 

\subsection{Vicinity of $u=-1/3$}

   To consider the approach to the point $u=-1/3$ from within the complex
extension of the FM phase, we use eq. (\ref{uform}) with $\epsilon \to 0$ and
hence (\ref{klim}).  We let $k_< \equiv i\kappa/\kappa'$, where 
$\kappa'$ is the complementary elliptic modulus satisfying
$\kappa^2+\kappa'^2=1$.  It follows that as $u \to -1/3$, $\kappa =
\epsilon/(4+\epsilon) \to 0$.  Using the identity \cite{gr} 
\beq
K(i\kappa/\kappa') = \kappa' K(\kappa)
\label{kidentity}
\eeq
we find that the term involving $K(k_<)$ in eq. (\ref{ctri}) approaches a
constant times $(1+3u)^{-1/2}$.  In the second term, one factor of
$(1+3u)^{-1/2}$ comes from the $k_<$ while another comes from the elliptic
integral $E(k_<)$, so that this term diverges like $(1+3u)^{-1}$.  This is the
leading divergence in $C$, so we thus obtain the exact result that as 
$u \to -1/3$ from within 
the complex FM phase, the critical exponent for the specific heat is 
\beq
\alpha_e' = 1
\label{alphae}
\eeq
To our knowledge, this is the first time an algebraic power has been found for
the specific heat critical exponent at a singular point in a 2D Ising model. 
 For the critical amplitude, we calculate (taking $\phi=0$ in eq. 
(\ref{uform}))
\beq
k_B^{-1} K^{-2} C \to  -\frac{2(3)^{3/2}}{\pi}|1+3u|^{-1}
\label{cue}
\eeq
The infinite set 
of $K$ values corresponding to the point $u=-1/3$ is 
\beq
K = \frac{1}{4}\ln 3 - \frac{(2n+1)i\pi}{4}
\label{ke}
\eeq
where $n \in Z$. 
\subsection{Vicinity of $u=-1$}

   We first observe that as $u \to -1$ from within the complex FM phase, 
$k_< \to -1$.  It follows that in this case the term involving $E(k_<)$ in
eq. (\ref{ctri}) is finite while the term involving $K(k_<)$ diverges
logarithmically, so that at this point $u_s=-1$, 
\beq
\alpha_s' = 0  \qquad (log \ div) 
\label{alphas}
\eeq
Thus, the divergence in the specific heat at $u=-1$ on the triangular lattice 
is of the same logarithmic type that it is \cite{egj,chisq} at $u=-1$ on the
square lattice, and the same as it is at the respective physical critical
points on both the square and triangular (as well as honeycomb) lattices. 
 For the critical amplitude, using the Taylor series expansion of $k_<$, as a
function of $u$, near $u=-1$, 
\beq
k_< = -1 - 2^{-3}(1+u)^3 + O((1+u)^4)
\label{klnearus}
\eeq
we calculate 
\beq
k_B^{-1}K^{-2}C \to \frac{12 i}{\pi}\ln |1+u| 
\qquad as \quad u \to -1
\label{cs}
\eeq
The infinite set of $K$ values corresponding to this point was given in
eq. (3.1.8) of paper I: $K = -i\pi/4 + n i \pi/2$ where $n \in Z$.  

   One can also consider the approach to $u=-1$ from within the
complex-temperature extension of the symmetric, PM phase (from the upper left
or lower left in Fig. 1(a)).  Using the exact expression for the specific heat
applicable in the symmetric phase \cite{ftri,domb1}, we find the same
logarithmic divergence in $C$, so that the corresponding exponent is 
\beq
\alpha_s=0 \qquad (log \ div)
\label{alphaspm}
\eeq

\section{Behaviour of $\chi$ at the Boundary of the PM Phase}

    For the square lattice, we have proved \cite{ms,chisq} that the
susceptibility has at most finite singularities on the boundary of the complex
 FM phase, aside from the physical critical point and the singular point at
$u=-1$.  We expect a similar result to hold for the triangular lattice,
although to show this with complete rigour, it would be desirable to perform 
an analysis of the asymptotic behaviour of the general connected 2-spin 
correlation function $<\sigma_{0,0}\sigma_{m,n}>$, as $r=(m^2+n^2)^{1/2} \to 
\infty$, which has not, to our knowledge, yet been done. 

   We have carried out a study of the high-temperature expansions for
$\bar\chi$ on the triangular lattice in order to search for 
complex-temperature singularities.  Since these singularities are presumably 
finite, we used the method of differential approximants, which is capable 
of representing such a finite singularity in a function even in the presence 
of an analytic background term.  The high-temperature series for $\bar\chi$ on
the triangular lattice was calculated in 1971 to $O(v^{16})$ \cite{high}.  We
have checked and found that apparently it has not been calculated to higher
order since \cite{gaunt}.   Mainly because of the shortness of the series, our 
study did not yield a definite value for the exponent $\gamma_s$ at the points
$v = \pm i$ corresponding to the point $u=-1$ as approached from within the
complex PM phase.  This might be possible with a substantially longer
high-temperature series. 

\section{Conclusions}

    In this paper we have investigated the complex-temperature singularities 
of the (zero-field) susceptibility $\bar\chi$ for the Ising model on the 
triangular lattice.  As part of this study, from an analysis of the 
low-temperature series using differential approximants, we have obtained
evidence which agrees with and strengthens the earlier conclusion by 
Guttmann \cite{g75} from ratio and dlog Pad\'{e} methods that $\bar\chi$ 
diverges like $|1-u/u_e|^{-\gamma_e'}$ at the endpoint $u_e=-1/3$ with 
exponent $\gamma_e'=5/4$ and have obtained a somewhat more precise 
determination of the critical amplitude, again in agreement with
Ref. \cite{g75}.  We also discussed how this
behaviour of $\bar\chi$ is related to the unusual feature that the spontaneous
magnetisation diverges at this point, and some implications of this.  From
exact results, we extracted the specific heat critical exponents at $u=-1/3$
and $u=-1$, which are, respectively, $\alpha_e'=1$ and $\alpha_s=0$ (log
divergence).  Finally, we have commented on non-analyticities of $\bar\chi$ 
which could occur as one approaches the boundary of the symmetric phase from 
within that phase. 
\vspace{4mm}

    One of us (RS) would like to thank Prof. David Gaunt and Prof. Tony
Guttmann for information about the current status of the series expansions 
for the triangular lattice.  RS would also like to thank Prof. Tony Guttmann
for kindly informing us of several of his early works on complex-temperature 
singularities and for discussions of these works. 
This research was supported in part by the NSF grant PHY-93-09888.

\vfill
\eject

\begin{center}
{\bf Figure Caption}
\end{center}

\vspace{5mm}

 Fig. 1. \  Phases and associated boundaries in the complex variables (a) $u$,
(b) $z$, and (c) $v$, as defined in eqs. (\ref{v})-(\ref{u}).  In Fig. 1(a),
the locus of points is given by the union of eqs. (\ref{ucircle}) and
(\ref{usegment}) in the text.  The intersections of the circle with the real
$u$ axis occur at $u_c=1/3$ and $u_s=-1$, the latter of which is also its
intersection with the semi-infinite line segment (\ref{usegment}) extending
from the point $u=-1/3$, denoted $u_e$ in the text, to $u=-\infty$ along the 
negative real axis.  In Fig. 1(b), the intersection of the outer boundary of
the complex FM phase with the real $z$ axis consists of the two points $\pm z_c
= \pm 1/\sqrt{3}$.  The two vertical line segments 
extend, respectively, from $\pm i/\sqrt{3}$ to $\pm i \infty$ along the 
positive (negative) imaginary axis. In Fig. 1(c), the intersections of the
curves with the real $v$ axis occur, from left to right, at $v=-1$, $v=v_c =
2-\sqrt{3}$, and $v=v_c^{-1}=2+\sqrt{3}$.  The endpoints of the arcs
occur at $v = \exp(\pm i\pi/3)$.  
\vfill
\eject

\end{document}